# Numerical Study on the Potential Enhancement of Organic Terahertz Sources through Tilted Pulse Front Pumping

Gyula Polónyi[1,2,✉], Zoltán Tibai[2], György Tóth[2], Gergő Krizsán[1,2] and János Hebling[1,2,3]

[1]*HUN-REN–PTE High-Field THz Research Group, Pécs, 7624, Hungary*
[2]*University of Pécs, Physics Department, Pécs, 7624, Hungary*
[3]*University of Pécs, Szentágothai Research Centre, Pécs, 7624, Hungary*
✉polonyi@fizika.ttk.pte.hu

**Abstract**

Organic crystals offer promising potential for THz generation, but face limitations in wavelength tunability and damage threshold. By applying tilted pulse front pumping to organic crystals an additional degree of freedom is introduced into the pumping conditions enabling a wider range of pumping wavelengths without compromising phase matching. Additionally, the lifespan of organic materials can be extended by using longer pumping wavelength and eliminate lower-order multi-photon absorption, allowing for higher pumping intensity without significant free carrier absorption, thus increasing the damage threshold. Simulations predict significant improvement for four out of six investigated crystals when tilted pulse front pumping is applied. By using volume phase holographic grating one can achieve pulse front tilt in organic crystals in collinear geometry with high diffraction efficiency. Design parameters are also presented.

**Keywords:** THz generation, organic crystals, tilted pulse front pumping

**Introduction**

High-field terahertz (THz) pulses in the 0.2 - 10 THz regime are pioneering tools for nonlinear THz optics, enabling the manipulation and control of material properties such as the ultrafast transformation of insulators into conductors [1], the controlled modulation of phase transitions and crystallization in chalcogenide crystals under the influence of intense THz fields [2], and even the acceleration of particles using table-top systems [3]. Optical rectification (OR) in nonlinear materials (ferroelectrics, semiconductors or organic crystals) provides THz pulses with high-fields necessary for driving the above applications.

Lithium niobate (LN) has emerged as a prominent nonlinear material for high-field, high-energy experiments, particularly when combined with the tilted pulse front pumping (TPFP) technique [4], which remarkable increased the conversion efficiency by 5 orders of magnitude [5]. TPFP has also found success in semiconductors, achieving an outstanding 220-fold enhancement in conversion efficiency in ZnTe by increasing the pumping wavelength from 800 nm to 1.7 $\mu$m [6]. This enhancement arises from a significant reduction in multi-photon absorption (MPA), as demonstrated in numerous experiments [7–9] even for organic crystals as well [10]. Furthermore, TPFP offers the intriguing possibility of tuning and broadening the generated frequency spectrum through its influence on phase-matching required for OR [11], thereby opening new avenues for tailored THz pulse generation.

MV/cm peak electric fields with percentage-level conversion efficiency can be routinely generated in organic crystals like OH1, or DAST in collinear pumping geometry if the pumping conditions are optimal. However, their low damage threshold remains a significant drawback. While increasing the pump wavelength can mitigate this issue by eliminating the low-order MPA, but – as Novelli observed – this comes at the cost of a significant decrease in conversion efficiency when one increases the pumping wavelength from 1.2 to 2.5 $\mu$m [12]. This decline stems from the growing phase mismatch between the optical and THz pulses at longer wavelengths.

In this context, a new idea emerges: to combine these two individually successful concepts, TPFP and organic materials for the generation of extreme high-field THz pulses, for overcoming existing limitations and pushing the boundaries of THz science towards new frontiers of discovery and application.

To the best of our knowledge, this is the first investigation to numerically explore THz generation from organic crystals pumped at longer wavelengths using a TPFP scheme. We achieve this by evaluating the analytical solution of the nonlinear wave equation and compare the different nonlinear materials through their conversion efficiency, furthermore, we determine the optimal pulse front tilt angle at a relatively long pumping wavelength that eliminates the low-order MPA. For these tilt angles we design the parameters of a volume phase holographic grating (VPHG) to achieve diffraction efficiencies above 90% in collinear pumping scheme replacing the traditional TPFP geometry. The selected organic materials under investigation are DAST, DSTMS, HMQ-TMS, OH1, BNA and MNA crystals.

**Organic crystals**

Organic crystals are attractive as THz sources as they are the most efficient THz generator materials with extremely high conversion efficiencies exceeding a few percent [13–16], and even surpassing 5% [10]. They possess high hyperpolarizability [17] owing to their unique molecule structure, and provide tens of MV/cm peak electric fields with pulse energies close to 1 mJ [14] generating THz radiation in collinear geometry. Their main drawback, as it was mentioned, is their low damage threshold, typically around 20 mJ/cm$^2$ for fs pumping while semiconductors can sustain 60 to 220 mJ/cm$^2$ (ZnTe, GaP, GaAs), and inorganics, like LN, withstand even 2.5 J/cm$^2$ [18]. Additionally, organic crystals reveal crystal degradation over prolonged exposure [19]. While achieving optimal conversion efficiencies requires specific pump wavelengths [14,15], these may not be as readily available as common lasers like Ti:sapphire or Yb lasers. Despite these limitations, organic crystals remain constantly in the center of interest.

**Coherence length**

In OR, phase matching requires that the phase velocity of THz and the group velocity of the pump are equal, from which the equality of the refractive indices is obvious:

$$n_{THz} = n_g. \qquad (1)$$

Dispersion in the materials usually disrupts the phase matching condition, limiting it to propagation lengths where the difference between the phases of the pump pulse envelope and the THz wave remains less than π. This length, expressed through the refractive indices, is known as the coherence length ($L_{coh}$) and characterizes the quality of phase matching according to the following equation:

$$L_{coh} = \frac{\pi c}{\omega \cdot |n_{THz} - n_g|}. \qquad (2)$$

Figure 1 shows the calculated coherence lengths for the most used organic crystals. As dispersion determines the optimal pumping wavelength for the nonlinear materials, this (in many cases) also limits the convenient pump sources to exotic ones (like chrome-forsterite lasers [14,15]), where fs pulse duration with high energy and high repetition rate (> 1 kHz) is not available currently. According to the figure, in case of DAST and DSTMS the necessary pumping wavelength is in the range of 1200-1300 nm, which is not so convenient. BNA and MNA need to be pumped below 1 $\mu$m where the MPA is strong, while OH1 and HMQ-TMS better performs above 1.6 $\mu m$ and 1.5 $\mu$m, respectively. Thus, decoupling phase matching from pumping wavelength can be beneficial.

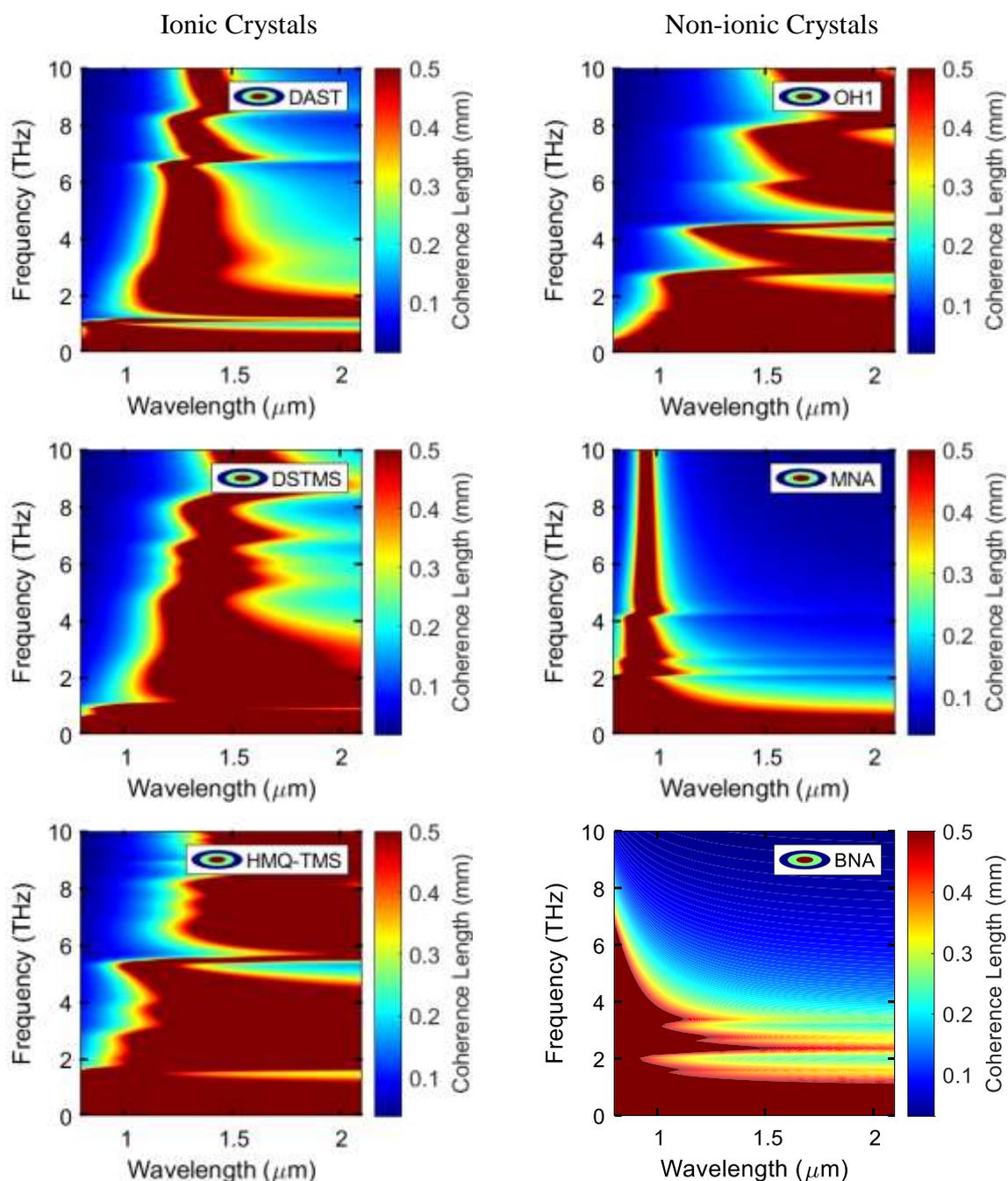

*Figure 1 – Coherence length of selected organic crystals (DAST, DSTMS, HMQ-TMS, OH1, MNA, BNA) for pumping between 0.8 to 2.1 µm (horizontal scale) and for THz frequencies between 0 to 10 THz (vertical scale).*

**Expected spectrum of the generated THz fields based on an analytical solution:**

In order to investigate the generated spectrums from OR in organic crystals at different pumping conditions, we use an analytical solution of the nonlinear wave equation in the plane-wave approximation considering transform-limited Gaussian pump pulses, both optical and THz-wave absorption, and non-perfect phase matching. However, it neglects nonlinear effects other than OR, and assumes non-depleted pump. The expression [17,20]:

$$E_{THz}(\omega, \lambda, L) = \frac{2 \cdot d_{THz}\omega^2 I(\omega)}{\left(\frac{\omega}{c}(n_{THz}+n_g)+i\left(\frac{\alpha_{THz}}{2}+\alpha\right)\right)n\epsilon_0 c^3} \cdot \frac{e^{i\left(\frac{\omega}{c}n_{THz}+\frac{i\alpha_{THz}}{2}\right)L} - e^{i\left(\frac{\omega}{c}n_g+i\alpha\right)L}}{\frac{\omega}{c}(n_{THz}-n_g)+i\left(\frac{\alpha_{THz}}{2}-\alpha\right)} \quad (3)$$

gives a good approximation of the generated electric field, where $\omega$ is the THz angular frequency, $\lambda$ is the pumping wavelength, $L$ is the crystal thickness, $d_{THz}$ is the nonlinear optical coefficient for THz generation, $I(\omega)$ is the spectral intensity of the pump, $n_{THz}$ and $\alpha_{THz}$ are the refractive index and the absorption coefficient in the THz range, $n_g$ and $\alpha$ are the group index and absorption coefficient in the optical range, $\epsilon_0$ is the vacuum permittivity and $c$ is the speed of light. The results for the six investigated organic crystals can be found in Fig. 4 (left panel). Even this simplified model aligns well with experimental observations regarding optimal pump wavelengths. For instance, HMQ-TMS exhibits the best performance when pumped around 1350 nm, as reported in [21]. For BNA, the shorter pumping wavelengths yielded higher conversion efficiencies [22]. Similarly, DSTMS achieves optimal results when pumped between 1200 and 1600 nm, rather than at longer wavelengths up to 2500 nm [12]. These findings not only validate the model's predictions, but also highlight the importance of considering material-specific optimal pumping wavelengths for achieving efficient THz generation in diverse organic materials.

**Effect of pump pulse duration**

To provide practical guidance for the most beneficial experimental designs, the setup should be tailored to the target application. Organic crystals excel in generating mid-range THz frequencies below 10 THz, thus the generated spectrum should ideally cover this range. Since the Fourier transform of the pump intensity, $I(\omega)$, appears in equation (3), the primary factor limiting the generated bandwidth is the pump pulse duration, which is inversely proportional with the spectral width. This spectral intensity can be written as $I(\omega) = \sqrt{2\pi}\tau I_0 A(\omega)$ [23], where $A(\omega)$ is the normalized Fourier spectrum $A(\omega) = \exp\left(-\frac{\tau^2\omega^2}{2}\right)$, which is plotted for different pump pulse durations (contour plot in Fig. 2). Additionally, the figure includes experimental data points indicating the achieved THz bandwidths for various pump pulse durations with specific organic crystals, demonstrating this trend. Based on this analysis, achieving an 8-10 THz broad spectrum necessitates a pump pulse duration of 100 fs or shorter.

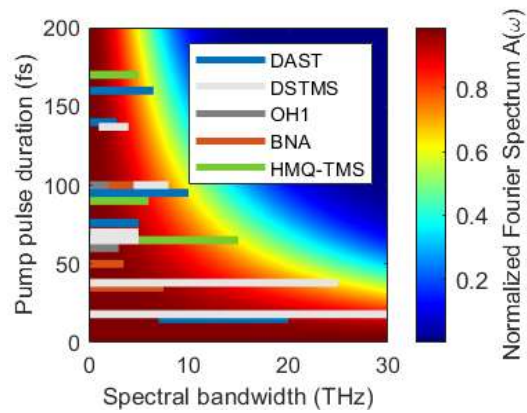

*Figure 2 – Contour plot of the normalized Fourier spectrum of the pump for various pump pulse durations. In addition, the experimental spectral widths are presented, which demonstrate good agreement with the theoretical predictions. References can be found in the supplementary material.*

**The proposed TPFP setup**

Tilted pulse front pumping was proposed [4] to achieve phase matching in materials like LN, which possesses high nonlinear coefficient, but lacks inherent matching between THz and optical refractive indices, due to a significantly smaller group index. When the pump pulse front is tilted by a dispersive optical element (prism or grating), the pump pulse and the THz beam propagate in different directions. Therefore, the projection of the pump group velocity onto the direction of the THz beam must be equal to the THz phase velocity. This will change the phase matching condition in Eq. (1) (expressed in terms of refractive indices):

$$n_{THz} = \frac{n_g}{\cos(\gamma)}, \quad (4)$$

where $\gamma$ represents the pulse-front-tilt (PFT) angle. To get rid of the effect of the resulting angular dispersion the dispersive optical element has to be imaged inside the crystal, which requires an imaging system that increases the complexity of the setup. Fortunately, for the selected organic crystals, the small needed PFT angles allow to circumvent this complexity by employing only a VPHG directly placed before the crystal. VPHGs have been recently proposed and demonstrated [24,25] to offer superior THz beam quality. They can achieve high diffraction efficiency at normal incidence in one specific diffraction order when inside them the planes with identical refractive indices are tilted (slanted) appropriately. This slant angle is denoted by $\varphi$ in Fig. 3. For efficient in-coupling, a refractive index matching liquid (RIML) or optical adhesive is proposed between the VPHG and the organic crystal. The grating period, ($d$) of the VPHG determines the diffraction angle within the nonlinear crystal (which must match the PFT angle) according to:

$$\sin(\gamma) = \frac{\lambda}{nd}, \quad (5)$$

where $\lambda$ is the pump wavelength in vacuum and $n$ is the refractive index. The generated THz pulse exits the crystal perpendicular to its back surface. For the simplest solution we assumed TE polarization (blue "x" in circles in Fig. 3). Here, the crystal must be oriented such that the crystal axis that exhibit the highest nonlinearity is parallel with the pump beam polarization. The parameters: slant angle, and the thickness of the VPHG are optimized for best diffraction efficiency in COMSOL Multiphysics software.

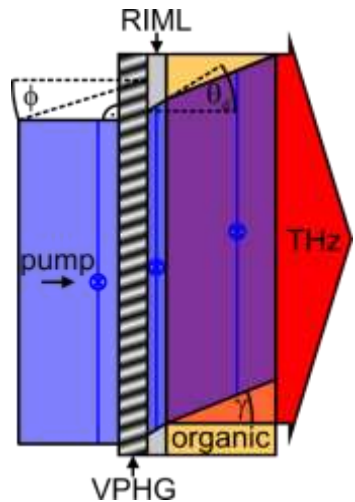

*Figure 3 – Proposed TPFP setup with VPHG before the organic crystal. The pump pulse enters perpendicularly into the VPHG, then becomes tilted by the diffraction on the slanted planes of angle φ. After that, the tilted pulse propagates through the refractive index matching liquid (RIML) and the organic crystal with the necessary pulse front tilt angle, γ. The generated THz pulse exits perpendicularly at the back surface of the crystal. The blue "x-in-circle" marks indicate the TE polarization. The setup ensures a convenient, collinear geometry for the applications.*

## Results & Discussion

Figure 4 presents the calculated electric fields using Eq. (3) with the following parameters: 10 GW/cm² intensity, 100 fs pump pulse duration, and 0.5 mm crystal thickness. The electric fields are plotted as a function of both the pump wavelength and the generated THz frequencies. The left panel shows the untilted case, while the right panel depicts the case with an optimally tilted pump pulse front (TPFP). In the latter case, $n_g$ was replaced by $n_g/\cos(\gamma)$ in Eq. (3) to account for the tilted pulse front. The characteristic vertical minima correspond to optical absorption of the pump, while the horizontal minima correspond to THz absorption of the crystal (see the absorption curves in the supplementary material). As can be seen, significant tilting angles (>10°) for most crystals (except OH1 and HMQ-TMS) induce noticeable changes in the contour plots.

Broadly speaking, the entire pattern undergoes a "tilt", wherein the high field regions at higher THz frequency distinctly shift towards longer wavelengths, while at lower THz frequency they experience a more gradual shift. This is because refractive indices are nearly equal at low THz frequencies, requiring minimal compensation. However, at higher frequencies, where the refractive index difference is larger, TPFP effectively compensates for the mismatch.

Application of TPFP leads to an increased area under the spectral curves, signifying higher pulse energy and conversion efficiency compared to the untilted case at the same pump wavelength. This is evident in Fig. 4 by comparing the vertical cross-sections at the dashed lines for both tilted and untilted cases. Due to the newly earned degree of freedom of selecting a pumping wavelength without compromising the velocity matching, it is possible to optimize for the highest area under the spectrum curves at a chosen wavelength (dashed line in Fig. 4).

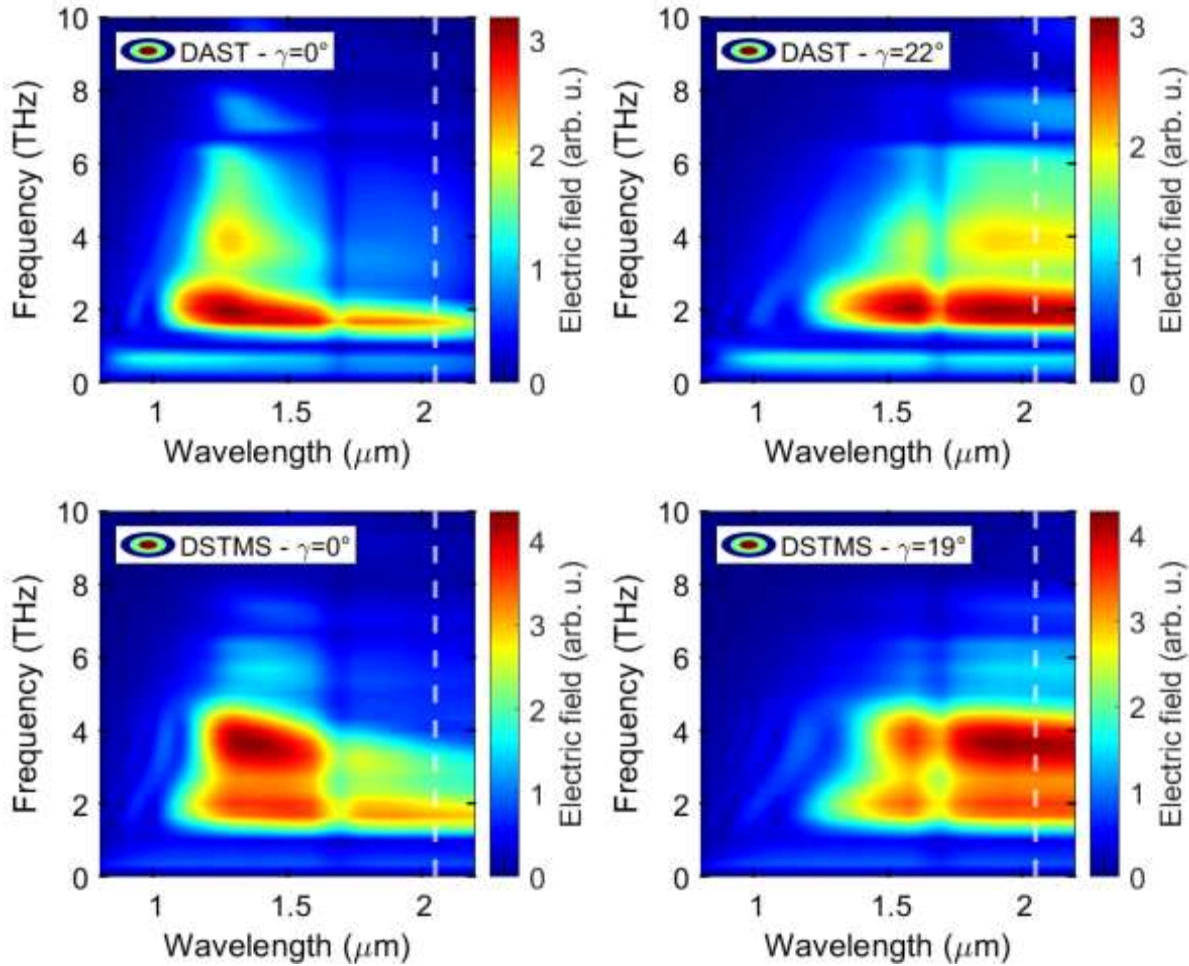

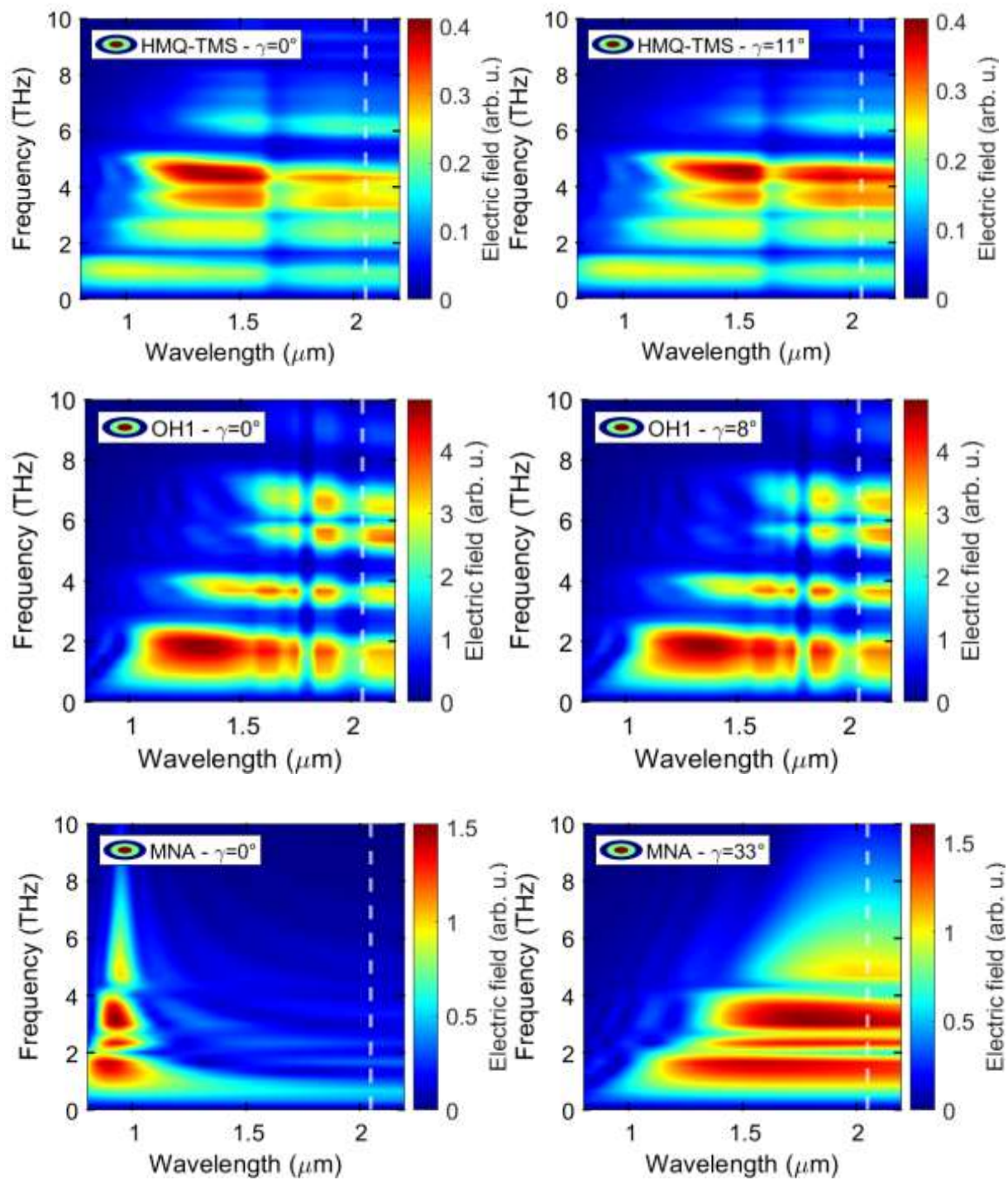

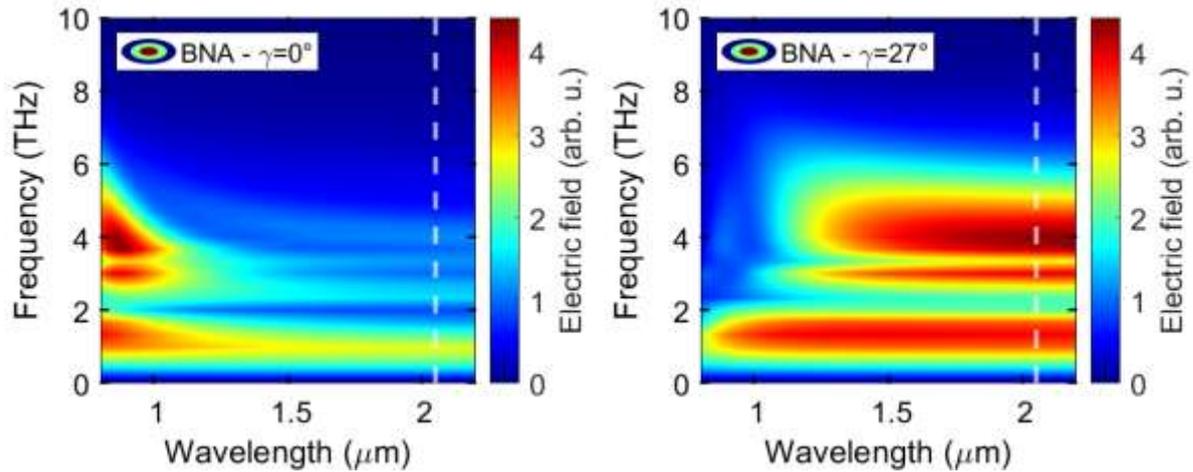

*Figure 4 – Evaluation of equation (3) presenting contour plot of the generated electric field between 0 to 10 THz for pumping wavelengths between 0.8 to 2.2 μm for untilted case (left panel) and for tilted case (right panel) with a tilt angle ($\gamma$) resulting the highest area under the spectrum curve at 2.05 μm pumping wavelength (vertical dashed line) for selected organic crystals.*

To demonstrate the potential benefits of using TPFP with organic crystals, we selected a pump wavelength (2050 nm) that satisfies three key criteria: it eliminates low-order MPA (2-3PA for DAST/DSTMS, 2-6PA for BNA/MNA), avoids strong optical absorption (refer to optical absorption curves in the supplementary material) and it is available with common holmium or thulium lasers. Fig. 5. presents the results of the optimization of the pulse front tilt angle. On the left side the ratio of the integrated spectrums of tilted and non-tilted cases versus the pulse front tilt angle are plotted, while on the right side the evaluated spectrums can be seen corresponding to the untilted case and at a pulse front tilt angle resulting in the highest $\frac{\int E_{titled} d\omega}{\int E_{untilted} d\omega}$ ratio for all the discussed crystals. The identical sharp minima in the blue and red curves, stem from characteristic THz absorption. In the regions at higher frequencies where the untilted spectra are smaller than the tilted ones, the discrepancy is due to phase mismatch, which can be corrected (in most of the cases) by TPFP.

The highest gain at 2050 nm pumping wavelength related to MNA crystal was with a 6-fold increment, followed by BNA (2.5x), DAST (2.3x), and DSTMS (1.7x). OH1 and HMQ-TMS appear insensitive to pulse front tilt because of their refractive index properties at these wavelengths as their group indices are already close to or exceed their THz refractive indices. Increasing the group index with pulse front tilt (division by $\cos(\gamma)$) further widens the gap, deteriorating phase matching. It is conceivable, however, that at even longer wavelengths (2.8 – 3 $\mu$m) TPFP could be of advance for these materials as well, but this requires additional studies and it may not be practical, if the conversion efficiency of the pumping OPA is also considered [26].

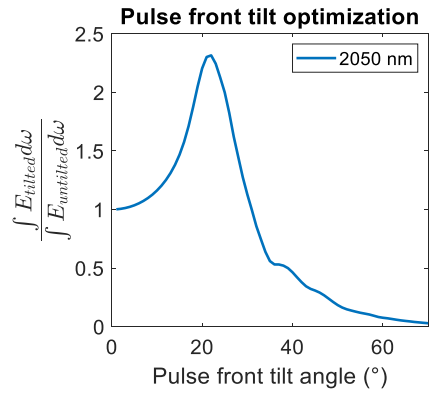 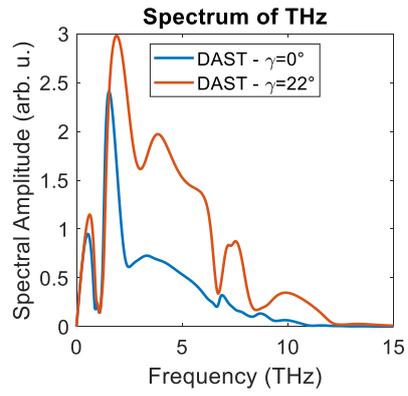

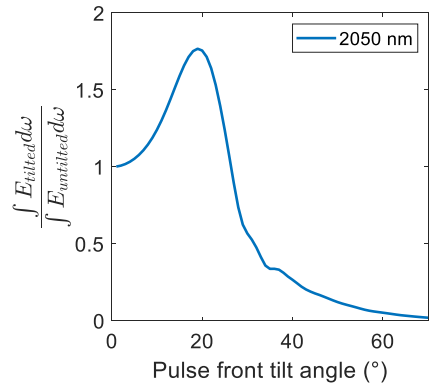 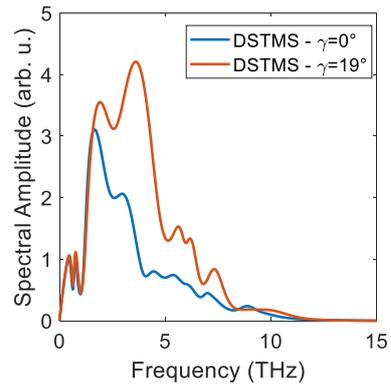

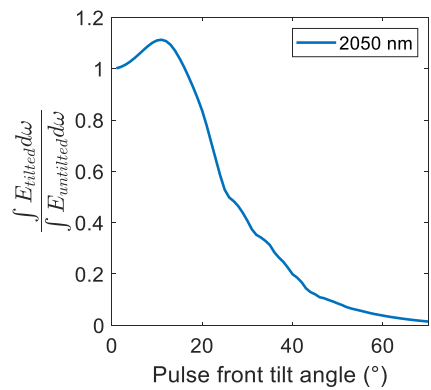 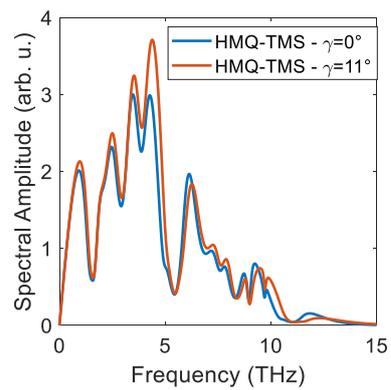

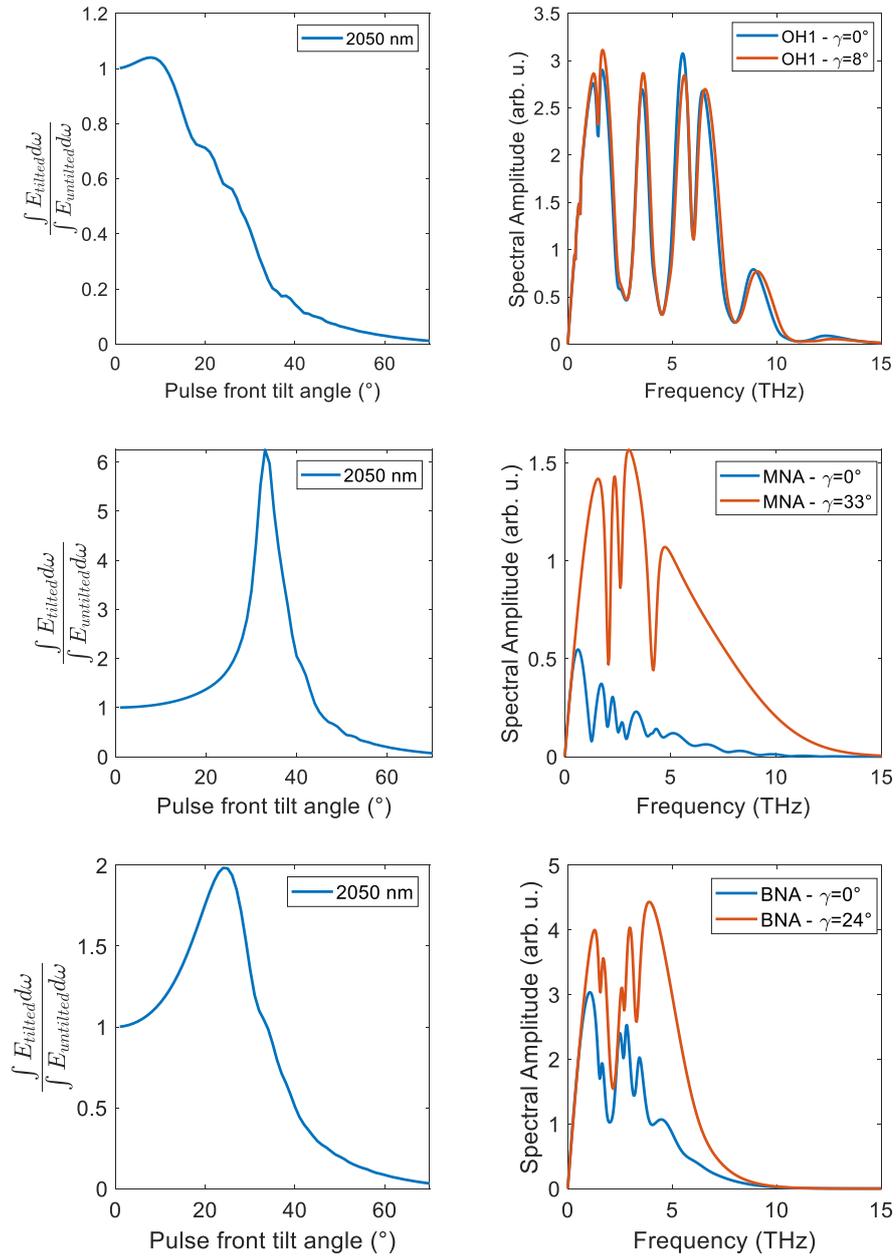

*Figure 5 – On the left panel the ratio of integrals of tilted-untilted spectrums vs tilting angle are presented, while the evaluated spectrums can be seen on the right panel corresponding to the normal incidence of the pump (blue) and the tilted case (red) with an angle resulting the highest ratio for all the discussed crystals.*

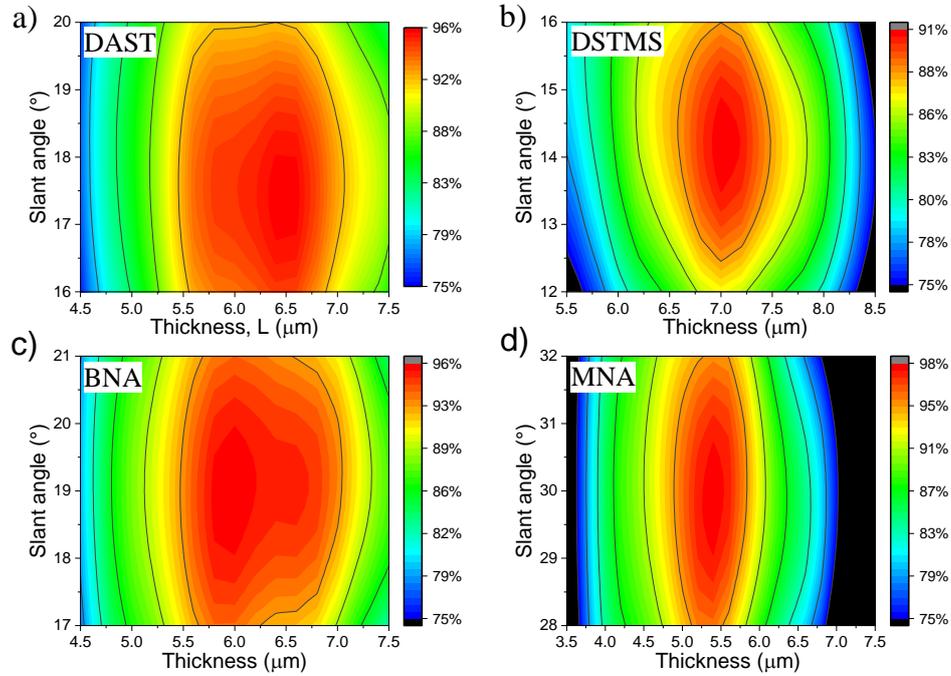

*Figure 6 – Simulated diffraction efficiencies as a function of slant angle and thickness of VPHGs suitable for selected organic crystals.*

As it was mentioned earlier, we proposed a very simple pulse front tilt geometry (see Fig. 3.), and for the four crystals which has proven to be the best the optimal VPHG parameters have been determined. According to our simulations carried out by COMSOL (see results in Figure 6.), diffraction efficiencies higher than 91% are expected for all the optimal VPHGs excluding OH1 and HMQ-TMS (Table 1.) with the calculated slant angles and VPHG thicknesses in TE mode. To compare the in-coupling efficiency of this setup with the scenario where the crystals are directly pumped without a VPHG, we calculated the transmission of the pump until it enters the crystal for both cases. In the first case, we considered the Fresnel loss from air to BK7, the diffraction efficiency of the grating, and the Fresnel loss between BK7 and the organic material (column before the last column). In the second case, only the Fresnel loss between air and the organic crystal was considered (last column). The results show that the difference is small, within 3%, but the gain in conversion efficiency is significantly higher. In some cases (DAST, MNA), the transmission itself is higher, resulting in a more advantageous THz source. Parameters for the calculation can be found in the supplementary material.

*Table 1. – Design parameters with the calculated VPHG diffraction efficiencies at 2050 nm. In case of slant angle and thickness, parameter ranges are given, within which the diffraction efficiency is not reduced by more than 5%.*

| Material | Dominant MPA order | PFT, $\gamma$ (°) | VPHG line density, $d$ (1/mm) | Slant angle, $\varphi$ (°) | Thickness ($\mu$m) | Diffraction efficiency (%) | Transmission (Fresnel losses + diff. eff.) (%) | Transmission w/o VPHG (%) |
|---|---|---|---|---|---|---|---|---|
| DAST | 4 | 22 | 385 | 15–20.2 | 5.4–7.2 | 96 | 89.6 | 87.3 |
| DSTMS | 4 | 19 | 326 | 6.3–7.7 | 12–16.6 | 91 | 85.3 | 88.1 |
| BNA | 7 | 27 | 392 | 16–21.9 | 5.4–7.1 | 96 | 91.5 | 92.3 |
| MNA | 7 | 33 | 526 | 27.7–32.2 | 4.8–5.9 | 98 | 91.9 | 89.2 |

**Conclusion**

TPFP offers a significant advantage for THz generation with organic crystals. By decoupling pump wavelength from phase matching, it allows selection from a broader range without compromising phase matching. This enables the use of longer pump wavelengths that eliminate lower-order MPA. Consequently, higher pump intensities can be employed without inducing strong free-carrier absorption. This translates to a crucial benefit: increased damage threshold and extended lifespan for organic crystals, which are typically limited by these factors. Numerical simulations demonstrated significant gains achievable with TPFP. Pumping DSTMS, DAST, BNA, and MNA crystals at 2050 nm with optimized tilt angles (19°, 22°, 27°, and 33°, respectively) resulted in conversion efficiency enhancements of 1.6x, 2.1x, 2.4x, and 6x compared to untilted pumping. To maintain a collinear geometry, a VPHG positioned before the crystals was proposed, with diffraction efficiencies exceeding 91% in all cases. OH1 and HMQ-TMS showed no improvement, in essence, with TPFP at 2050 nm due to near phase matching, but potentially better results might be achievable at even longer wavelengths where their group index is lower. Since VPHGs preserve the collinear geometry, they also facilitate the generation of radially polarized THz pulses, desirable for applications like particle acceleration. This can be achieved using a cascaded setup with 2-6 VPHG-crystal pairs rotated relative to each other and employing segmented pumping [27].

The TPFP technique can be used in combination with other damage threshold increasing methods like adding a high-thermal conductivity window to the input surface of the organic crystal [19,28] or cooling it down to cryogenic temperature [28]. The investigations can be continued with a more advanced model considering additional nonlinear effects like cascading, multi-photon absorption, free-carrier absorption and self-phase modulation, introduced in [26].

**Funding**

National Research, Development and Innovation Office (2018–1.2.1-NKP-2018–00009, 2018–2.1.5-NEMZ-2018–00003, 2018-1.2.1-NKP 2018-00010, TKP2021-EGA-17), Project 101046504 — TWAC, Funded by the European Union. Views and opinions expressed are however those of the author(s) only and do not necessarily reflect those of the European Union or the European Innovation Council and SMEs Executive Agency (EISMEA). Neither the European Union nor the granting authority can be held responsible for them. The project has been supported by Development and Innovation Fund of Hungary, financed under the TKP2021-EGA-17 funding scheme.

**Conflict of Interest**

All authors declare no competing interests.

**Data Availability**

Data underlying the results presented in this paper are not publicly available at this time but may be obtained from the authors upon reasonable request.